\documentclass[conference]{IEEEtran}
\IEEEoverridecommandlockouts
\usepackage{enumitem}
\usepackage{cite}
\usepackage{amsmath, amssymb}
\usepackage{xcolor}
\usepackage{algorithm}
\usepackage{algpseudocode}
\usepackage{amsthm}    
\usepackage{amsmath}
\usepackage{amssymb} 
\newtheorem{theorem}{Theorem}

\usepackage{tikz}
\usetikzlibrary{positioning, shapes}
\usepackage{mathtools} 
\usepackage{todonotes}

\newcommand{\R}{\mathbb{R}}
\newcommand{\No}{\mathbb{N}_0}
\usepackage{amsthm}  
\usepackage{amsmath}
\usepackage{amssymb}
\usepackage{xcolor}
\usepackage{graphicx} 
\usepackage{subcaption}
\usepackage[font=small]{caption} 
\def\BibTeX{{\rm B\kern-.05em{\sc i\kern-.025em b}\kern-.08em
    T\kern-.1667em\lower.7ex\hbox{E}\kern-.125emX}}
\usepackage[framemethod=TikZ]{mdframed}
\usepackage{mdframed}
\usepackage{algorithm}
\usepackage{algpseudocode}
\makeatletter
\renewcommand{\ALG@beginalgorithmic}{\setcounter{ALG@line}{0}}
\makeatother

\newcommand{\E}{\mathbb{E}}

\newcommand{\T}{{\tiny \intercal}}

\newcommand{\tr}{\mathrm{tr}}
\newcommand{\es}{e^{\rm s}}
\newcommand{\Jb}{\bar{J}}
\newcommand{\bt}[1]{\bar{\theta}_{#1}}


 
\begin{document}

\title{A Linear Programming Framework for Optimal Event-Triggered LQG Control}
\author{Zahra Hashemi and Dipankar Maity%
\thanks{
This research is supported by the National Science Foundation CAREER Award 2443349.}
\thanks{
The authors are with the Department of Electrical and Computer Engineering at the University of North Carolina at Charlotte, NC, USA, 28223. (e-mails: {\tt \{zahrahashemi1, dmaity\}@charlotte.edu}).}%
}

\maketitle

\begin{abstract}
This paper explores intelligent scheduling of sensor-to-controller communication in networked control systems, particularly when data transmission incurs a cost. While the optimal controller in a standard linear quadratic Gaussian (LQG) setup can be computed analytically, determining the optimal times to transmit sensor data remains computationally and analytically challenging. We show that, through reformulation and the introduction of auxiliary binary variables, the scheduling problem can be cast as a computationally efficient mixed-integer linear program (MILP). This formulation not only simplifies the analysis but also reveals structural insights and provides clear decision criteria at each step. Embedding the approach within a model predictive control ({\small MPC}) framework enables dynamic adaptation, and we prove that the resulting scheduler performs at least as well as any deterministic strategy (e.g., periodic strategy). Simulation results further demonstrate that our method consistently outperforms traditional periodic scheduling.
\end{abstract}


\section{Introduction}

Networked control systems ({\small NCS}) play a vital role in today’s cyber–physical systems, where sensors, actuators, and controllers interact through shared communication networks. This architecture offers significant advantages in terms of flexibility and scalability. However, it also presents key challenges, particularly due to limited communication resources. Continuously transmitting sensor data at every time step can be inefficient or even impractical, especially under  energy \cite{heemels2012introduction, maity2019optimal,hashemi2016energy}, bandwidth \cite{  afshari2024communication, maity2021optimal, kostina2017rate}, and latency constraints \cite{   maity2018optimal, maity2021optimalb}.

A substantial body of research has explored event- and self-triggered, periodic, and sampled-data control, as a way to balance control performance with communication efficiency \cite{aastrom1997fundamental}.
Early approaches focused on deterministic triggering rules that guarantee stability by transmitting whenever a predefined state or error threshold is exceeded \cite{rabi2012adaptive,imer2010optimal,lipsa2011remote, maity2015eventa}. More recent work has shifted toward stochastic and optimization-based methods, where transmission policies are co-designed with control laws to minimize objective functions that explicitly penalize communication~\cite{molin2009lqg,molin2012optimality, maity2020minimal, mamduhi2025network, mamduhi2023regret, maity2023regret}.

Even with recent progress, there is still no widely applicable and computationally efficient algorithm for tackling the scheduling problem to find the optimal communication times. The main challenge lies in the bilinear nature of the error covariance recursion, combined with a decentralized setup where the scheduler and controller have access to different information. This mismatch makes classical dynamic programming techniques ineffective and results in a complex, nonconvex stochastic optimization problem with high dimensionality \cite{molin2009lqg}.

Several alternative approaches have been explored. Among these, \textit{value-of-information} (VoI) based approaches attempt to quantify the benefit of each transmission, but depend on dynamic programming in continuous state-space, which quickly becomes impractical as system size grows \cite{soleymani2021value,soleymani2022value}. In continuous-time settings, partial differential equations and Hamilton-Jacobi-Bellman based formulations have been proposed \cite{thelander2020lqg}, but solving these equations is computationally intensive, even for relatively small systems. These limitations highlight the need for reformulations that retain optimality while offering better scalability.
A recent study has also considered deep reinforcement learning based methods to overcome this analytical and computational intractability \cite{aggarwal2025interq}. 

To tackle these issues, we propose a scheduler-centric reformulation that retains optimality while enabling efficient computation. The key idea is to eliminate matrix-valued decision variables arising from the covariance recursion and instead represent the nonconvex switching behavior using auxiliary binary monomials. This leads to an equivalent mixed-integer linear program (MILP) with a linear objective and linear constraints—making it solvable with modern integer programming tools.

This paper makes two primary contributions. First, we present an exact reformulation of the stochastic scheduling problem with bilinear error dynamics. Building on the certainty-equivalent reduction introduced by Molin \emph{et. al} \cite{molin2009lqg,molin2012optimality}, we show that the problem can be transformed into a deterministic mixed-integer nonlinear program (MINP). By unfolding the covariance recursion, we eliminate matrix-valued decision variables, reducing the problem to an unconstrained MINP that depends solely on binary scheduling decisions. Through the introduction of binary monomials and the application of McCormick relaxation, we construct a MILP that is exactly equivalent to the original formulation. This reformulation enables efficient computation.
Second, we develop a real-time method for transmission decisions using one-step send/skip certificates. By comparing local and future Gramians, the certificate determines whether sending or skipping is optimal at each step, often allowing the scheduler to avoid solving the MILP. This mechanism integrates efficiently into a model predictive control ({\small MPC}) framework and enables fast, adaptive scheduling. We prove that the resulting {\small MPC}-based scheduler strictly outperforms all deterministic causal policies, including periodic strategies. In numerical simulations across varying system sizes, our MILP reformulation achieves speedups of up to five orders of magnitude—reducing optimization time by factors of $10^5$ compared to solving the original MINP directly. Moreover, in a double-integrator case study, we demonstrate that the {\small MPC} scheduler consistently achieves lower control cost with fewer transmissions than baseline methods.

The paper is organized as follows. We describe the problem formulation in Section~\ref{sec:problem} and the optimal controller in Section~\ref{sec:control}. 
The event-triggered communication policy and the linear programming reduction is presented in \ref{sec:communication}. 
The efficiency and optimality certificates of the proposed method are addressed in Section~\ref{sec:MPC_certificates}. Section~\ref{sec:case-study} reports the simulation results and we conclude the work in Section~VIII. 
 
\textit{Notation:} Let $\R$ and $\No$ denote the set of real numbers and non-negative integers, respectively.
All vectors and matrices are real valued with compatible dimensions. For a symmetric matrix $Q$, define the weighted norm $\|x\|_Q^2:=x^\T Qx$. We use the Loewner order: $A\preceq B$ means $B-A\succeq0$; in particular $Q\succeq0$ denotes positive semidefinite matrices. The operator $\operatorname{tr}(\cdot)$ denotes the trace. Expectation and probability are denoted by $\mathbb{E}[\cdot]$ and $\Pr(\cdot)$, respectively. Unless stated otherwise, expectations are taken with respect to the initial state $x_0$ and the process-noise sequence $\{w_k\}_{k\in\No}$. Conditional expectations use the controller’s information set, written as $\mathbb{E}[\cdot \mid \mathcal{Z}_k]$.

\section{Problem Description} \label{sec:problem}
Let us consider a discrete-time stochastic networked control system following the dynamics:
{\small
\begin{equation}
    x_{k+1} = A x_k + B u_k + w_k,
    \label{eq:discrete-system}
\end{equation}
\begin{equation}
    z_{k} = \begin{cases}
        x_k \qquad &\theta_k = 1,\\
        \emptyset \qquad & \theta_k = 0,
    \end{cases} 
    \label{eq:measurement}
\end{equation}
}where \( A \in \mathbb{R}^{n \times n} \) and \( B \in \mathbb{R}^{n \times m} \) are constant system matrices.
The state vector \( x_k \in \mathbb{R}^n \) and control input \( u_k \in \mathbb{R}^m \) are updated at each discrete timestep \( k \). 
In \eqref{eq:measurement}, the variable $z_k$ denotes the measurement \emph{received} by the controller at time $k$, which depends on the communication decision variable $\theta_k \in \{0,1\}$. 
The sensor measurement is $x_k$, which may or may not be transmitted, depending on the communication decision. 
This formulation enables event-triggered communication, where the sensor transmits measurements selectively, conditioned on the occurrence of a triggering event.

The initial state \( x_0 \) is modeled as a random variable with finite mean and covariance. The process noise \( w_k \in \R^n \) is an independent and identically distributed (i.i.d.) sequence of zero-mean random variables with finite covariance $\Sigma^w$. Additionally, \( x_0 \) and the noise variables \( w_k \) are assumed to be statistically independent for every \( k \).

\begin{figure}
\centering
\resizebox{0.55\columnwidth}{!}{%
\begin{tikzpicture}[auto, node distance=1.1cm, >=latex, thick]

  \node[draw, rectangle] (C) { Controller};
  \node[draw, rectangle, right=of C] (A) {Plant};
  \node[draw, circle, right=of A, minimum size=8mm, inner sep=0pt, color=white] (mult) {};
\node[draw, rectangle, above=of mult, yshift=-5mm] (Scheduler) {Scheduler};

  \node[draw, very thick,
        cloud, cloud puffs=12, cloud puff arc=120, cloud ignores aspect,
        minimum width=3cm, minimum height=1.3cm,
        below=1cm of A] (Network) {\textbf{Network}};

  \draw[->, very thick] (C) -- node[midway,above] {$u_k$} (A);
  \draw[-, very thick] (A) -- node[midway,above] {$x_k$} +(1.5,0) 
            coordinate[pos=0.25] (branchX) -- +(2.5,-.5);
  \draw[->, very thick] (branchX) |- (Scheduler.west);
  \draw[->, very thick] (Scheduler.south) -- node[right] {$\theta_k$} (mult.north);
  \draw[->, very thick] (mult) -- node[midway,above] {$z_k$}  +(1,0) |- (Network.east);

  \draw[->, very thick] (Network.west) -| ([yshift=0cm, xshift=-4.5mm] C.south west) |- (C.west);

\end{tikzpicture}
}
\caption{\footnotesize System architecture illustrating the interaction among the plant, scheduler, network, and controller. At each time step \( k \), the scheduler selects~a binary transmission decision \( \theta_k \). Based on this decision, the corresponding measurement $z_k \in \{x_k, \varnothing\}$ is transmitted to the controller which computes the input \( u_k \)}.
\label{fig:system-configuration} \vspace{-10pt}
\end{figure}
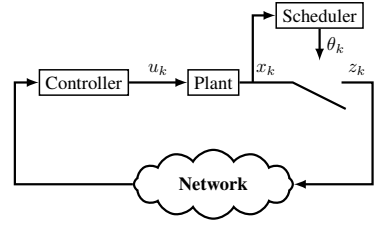

To formalize the controller’s access to measurement data over time, we introduce the \emph{information set} available to the controller at each time step. 
We assume that the controller also remembers its past control actions. 
Let $\mathcal{Z}_k = \{ z_{0:k}, \theta_{0:k}, u_{0:k-1} \}$ denote the information available to the controller at time step $k \ge 1$, with the initial information set given by $\mathcal{Z}_0 = \{ z_{0}, \theta_{0} \}$. 
At the next timestep, the information set is updated as:

{\small
\begin{equation}
\mathcal{Z}_{k+1} = \mathcal{Z}_k \cup \{ z_{k+1}, \theta_{k+1}, u_k \}.
\end{equation}
}We consider the system’s evolution over a finite horizon of \( T \) steps. Fig.~\ref{fig:system-configuration} depicts the overall system architecture, which consists of the following components:
\begin{itemize}[leftmargin=*]

  \item \textbf{Scheduler:} Observes the state \( x_k \) at each discrete time step \( k \), and makes a binary decision \( \theta_k \in \{0,1\} \) indicating whether the controller should receive the current measurement via the resource limited network.

  \item \textbf{Network:} Transmits the packet \( \{z_k, \theta_k\} \) from the plant to the controller. For simplicity, we assume an ideal network with no delays, quantization, or packet loss.

\item \textbf{Controller:} Based on the information set \( \mathcal{Z}_k \), the controller computes the control input \( u_k \) and applies it to the plant.

\item \textbf{Plant:} Generates the next state \( x_{k+1} \) according to the dynamics in~\eqref{eq:discrete-system}.

\end{itemize}
The primary objective is to jointly design a control policy \( \mu \) and a communication scheduling policy \( \Theta =\theta_{0:T-1} \) to minimize the expected cost functional:
{\small
\begin{equation}
J(\mu, \Theta) = \mathbb{E} \left[ \sum_{k=0}^{T-1} \left( \|x_k\|^2_Q + \|u_k\|^2_R + \lambda \theta_k \right) + \|x_T\|^2_{Q_T} \right],
\label{eq:discrete-cost-function}
\end{equation}
}where the expectation \( \mathbb{E}[\cdot] \) is taken with respect to the stochasticity in the initial state, process noise, and—if randomized policies are considered—the scheduling and control decisions. 
Here, \( Q \succeq 0 \) and \( Q_T \succeq 0 \) are state weighting matrices, \( R \succ 0 \) is the control weighting matrix, and \( \lambda > 0 \) is a designer-specified regularization parameter that penalizes communication. 
The term \( \sum_{k=0}^{T-1} \theta_k \) quantifies the total number of transmissions, directly impacting network usage. 
Therefore, the cost function \eqref{eq:discrete-cost-function} captures a trade-off between control performance and communication effort, accumulated over the horizon \([0,T]\).

Formally, we aim to perform the following joint optimization over the control and communication variables
{\small
\begin{align*}
    \min_{\mu,\Theta} J(\mu,\Theta).
\end{align*}
}
\section{Optimal Controller} \label{sec:control}

Solving the joint optimization problem of control law and event-triggering schedule in~\eqref{eq:discrete-cost-function} is a central challenge in event-triggered control. 
The fact that the scheduler and controller operate with different information sets introduces a decentralized information structure. As a result, classical dynamic programming techniques are not directly applicable, prompting the development of alternative solution methods. This challenge—optimizing control under limited and decentralized information—has been widely studied, with several formulations and solution strategies proposed in the literature. Recent work shows that the problem can sometimes be decomposed into simpler subproblems, making it more tractable. Notably, some parts of the formulation resemble the standard LQR problem, which can be efficiently solved using the Riccati equation.

We begin by examining the event-triggered control considerted by Molin \emph{et al.}~\cite{molin2009lqg,molin2012optimality}, which proved that the optimal control policy in this discrete-time setting has the form:

{\small
\begin{equation}
u_k = -L_k \mathbb{E}[x_k \mid \mathcal{Z}_k], \quad k \in \{0, \ldots, T-1\},
\label{eq:molin-control}
\end{equation}
}where recall that {\small\(\mathcal{Z}_k\)} denotes the information set available to the controller up to time {\small\(k\)}. 
The control gain {\small\(L_k\)} and the associated Riccati recursion are given by:
{\small
\begin{subequations}
\begin{align}
L_k &= S_k^{-1} B^\T P_{k+1} A, \\
S_k &= R + B^\T P_{k+1} B, \\
P_k &= A^\T P_{k+1} A + Q - A^\T P_{k+1} B S_k^{-1} B^\T P_{k+1} A,
\end{align}
\end{subequations}
}with terminal condition {\small\(P_T = Q_T\)}, and where {\small\(P_k \in \mathbb{R}^{n \times n}\)} is nonnegative definite for all {\small\(k \in \{0, \ldots, T\}\)}. By defining, $\hat{x}_k := \mathbb{E}[x_k \mid \mathcal{Z}_k]$ we may verify \cite{molin2012optimality}

{\small
\begin{align*}
    \hat{x}_k = \begin{cases}
        x_k & \theta_k = 1, \\
        A \hat{x}_{k-1} + B u_{k-1} & \theta_k =0.
    \end{cases}
\end{align*}
}As further shown in~\cite{molin2012optimality}, substituting the optimal control law \eqref{eq:molin-control} into the cost functional reduces the problem to minimizing a new cost that depends solely on the transmission decision vector \(\Theta\):

{\small
\begin{equation}
J(\mu^*, \Theta) = J_{\text{const}} + \mathbb{E} \bigg[ \!\sum_{k=0}^{T-1} \!\! \|e_k\|^2_{L_k^\T S_k L_k} \bigg] + \lambda \mathbb{E} \left[ \sum_{k=0}^{T-1} \theta_k \right],
\label{eq:cost-function2}
\end{equation}
}where \( e_k = x_k - \mathbb{E}[x_k \mid \mathcal{Z}_k] \) denotes the state estimation error at time \(k\). Here, \(J_{\mathrm{const}}\coloneqq
\mathbb{E}\!\big[x_0^{\T} P_0 x_0\big]
+\sum_{k=0}^{N-1}\mathbb{E}\!\big[w_k^{\T} P_{k+1} w_k\big]\)
denotes the part of the cost that depends only on the initial state \(x_0\) and
the process noise sequence, and is therefore independent of the transmission
vector \(\Theta\)~\cite{molin2012optimality}.
 Following the analysis in~\cite[Theorem~1]{molin2009lqg}, the error sequence \(\{ e_k \}_{k \in \No}\) evolves recursively~as:
 
 {\small
\begin{equation}
e_{k+1} = \bar{\theta}_{k+1} (A e_k + w_k),
\label{eq:error-dynamics}
\end{equation}
}with initial condition \( e_{-1} = 0 \), where \( \bar{\theta}_{k+1} := 1 - \theta_{k+1} \). That is, the estimation error resets to zero whenever a fresh measurement is received (\( \theta_{k+1} = 1 \)), and otherwise propagates forward under the system dynamics with additive process noise. This formulation serves as the foundation for the event-triggered estimator used in~\eqref{eq:cost-function2}. For further details and theoretical foundations, we refer the reader to~\cite{molin2009lqg}.

\section{Optimal Communication Protocol} \label{sec:communication}
In this section, we investigate the optimal scheduling policy, which was left out in \cite{molin2009lqg, molin2012optimality}, and still remains to be an intractable problem. 
To this end, we define the error on the scheduler's side at time \(k\) as:
{\small
\begin{align*}
    \es_k &:= x_k - A \hat{x}_{k-1} - B u_{k-1} = A e_{k-1} + w_{k-1}\\
    \es_0 &:= x_0 -\E[x_0].
\end{align*}
}One may verify that the estimation error at the controller side follows $e_k = \bt{k} \es_k$ for all $k$. 
At any given $k$, the scheduler optimizes for the remaining horizon:

{\small
\begin{align} \label{eq:MPC_optt}
    \min_{\bt{k:(T-1)}} \quad & \E\left[\sum_{t=k}^{T-1} \|e_t\|^2_{\Gamma_t} + \lambda (1-\bt{t})\right] \nonumber \\
    \text{subject to} \quad & e_{t+1} = \bt{t+1}(A e_t + w_t) \\
    & e_k = \bt{k}\es_k. \nonumber    
\end{align}
}Here, $\Gamma_t := L_t^\T S_t L_t$ denotes the weighting matrix appearing in the cost functional~\eqref{eq:cost-function2}.
Notice that the scheduler error $\es_k$ affects the scheduling policy for the remaining horizon $[k, (T-1)]$. 
Given $\es_k$, we find the optimal scheduling {\small$\Theta_k^*(\es_k) := \{\bt{k|k}^*, $ $ \bt{k+1|k}^*,  \ldots, \bt{T-1|k}^*\} $ }and apply the first one, i.e., $\bt{k|k}^*$, and then repeat the same process at time $k+1$ with the new realized error $\es_{k+1}$.

To develop an efficient algorithm for solving the stochastic optimization \eqref{eq:MPC_optt},  we proceed as follows. 
For any given decision roll-out $\Theta_k = \{\bt{k|k}, \ldots, \bt{T-1|k}\}$, we define $\Jb( \Theta_k, k |  \es_k)$ to be the objective value of the optimization \eqref{eq:MPC_optt}. 
That is, 

{\small
\begin{align}
    \Jb( \Theta_k, k |  \es_k) = \sum_{t=k}^{T-1} \tr(\Gamma_t \Sigma_t) + \lambda(1-\bt{t|k}),
\end{align}
}where {\small$\Sigma_t = \E[e_te_t^\T ~|~\Theta_k,\es_k]$} for all $t=k,\ldots,(T-1)$.
Consequently, 
{\small
\begin{align*}
    \Sigma_{t+1} &=  \E[e_{t+1}e_{t+1}^\T~|~\Theta_k,\es_k] \\
    & = \E[\bt{t+1|k}^2 (Ae_t + w_t)(Ae_t + w_t)^\T~|~\Theta_k,\es_k] \\
    &\overset{(*)}{=} \bt{t+1|k} \E[ Ae_t e_t^\T A^\T + Ae_t w_t^\T + w_t e_t^\T A^\T + w_t w_t^\T~|~\Theta_k,\es_k  ] \\
    & \overset{(\dagger)}{=} \bt{t+1|k} (A\Sigma_t A^\T + \Sigma^{w}),
\end{align*}
}where $(*)$ follows from the fact that {\small$\bt{t+1|k}^2 = \bt{t+1|k}$} and $\bt{t+1|k}$ is `deterministic' given $\Theta_k$, which allows us to take $\bt{t+1|k}$ out of the expectation.
Then, $(\dagger)$ follows from the definition of $\Sigma_t$ and the fact that $w_t$ has zero mean and covariance $\Sigma^w$ and is independent of $e_k^s$ and the choice of $\Theta_k$ for all $t$. 
Thus, the original optimization under stochastic dynamics can be reformulated as a deterministic \emph{mixed-integer nonlinear program } with bi-linear matrix equality constraints:
\begin{mdframed}[backgroundcolor=gray!10, roundcorner=10pt, innerleftmargin=10pt, innerrightmargin=10pt]
\small
\textbf{Matrix Recursion Form}
{\small
\begin{align} \label{eq:MPC_2}
\begin{split}
    \min_{~~~~~\Theta_k,\,\Sigma_{k:T-1}}  &\Jb(\Theta, k \mid e_k^s) \\
    \text{subject to} \quad & \Sigma_{t+1} = \bt{t+1|k} (A\Sigma_t A^\T + \Sigma^{w}), \\
    & \hfill \qquad \qquad \qquad \qquad  \forall ~t=k,\ldots,T-2, \\
    & \Sigma_k = \bt{k|k} e_k^s (e_k^s)^\T. 
\end{split}
\end{align}
}
\end{mdframed}
Even though the objective function is linear (both in $\Sigma_{k:T-1}$ and $\bt{k:T-1}$), the bi-linear matrix equalities pose a significant challenge in solving the optimization. 
These matrix variables also increase the dimensionality of the optimization problem---{\small$(T-1-k)(\tfrac{n(n-1)}{2}+1)$} variables for the $k$-th timestep optimization. 
The major bottleneck (i.e., the {\small$(T-1-k)\tfrac{n(n-1)}{2}$} part) comes from the matrix variables, which scales quadratically with the dimension of the state space $n$. 

\subsection{Mixed Integer Linear Program Reformulation}

To overcome the mentioned computational challenge, we reformulate this optimization problem by completely eliminating the matrix variables. Unfolding the bi-linear matrix dynamics yields the following closed-form expression:

{\small
\begin{align}
\Sigma_t
&= \left(\prod_{s=k}^{t}\bar{\theta}_{s|k}\right)
   A^{\,t-k}\,\Sigma_k^s\,A^{\,t-k^\T} \nonumber \\
&\quad + \sum_{\tau=k+1}^{t}
   \left(\prod_{s=\tau}^{t}\bar{\theta}_{s|k}\right)
   A^{\,t-\tau}\,\Sigma^{w}\,A^{\,t-\tau^\T}.
\label{eq:Sigma-sum}
\end{align}
}Let us further define 
{\small
\begin{align}
G_{t,\tau} &:= 
\begin{cases}
A^{\,t-k}\Sigma_k^s A^{\,t-k^\T}, & \tau=k, \\[2pt]
A^{\,t-\tau}\Sigma^{w}A^{\,t-\tau^\T}, & \tau\ge k+1,
\end{cases} \label{eq:def-K}\\[4pt]
g_{t,\tau} &:= \operatorname{tr}\!\big(\Gamma_t G_{t,\tau}\big). \label{eq:def-c}
\end{align}
}With these definitions, the closed-form covariance expression in \eqref{eq:Sigma-sum} 
can be rewritten as the scalar-weighted expansion:
{\small
\[
\Sigma_t = \sum_{\tau=k}^{t} \Big(\prod_{s=\tau}^{t} \bar\theta_{s|k}\Big) G_{t,\tau},
\]
}and the instantaneous stage cost simplifies to
{\small
\[
\operatorname{tr}(\Gamma_t \Sigma_t) 
= \sum_{\tau=k}^{t} \Big(\prod_{s=\tau}^{t} \bar\theta_{s|k}\Big) g_{t,m},
\]
}where all coefficients $g_{t,\tau}$ can be precomputed \textit{offline}. Upon substituting \eqref{eq:Sigma-sum} into the objective function, we~obtain an \textit{unconstrained mixed-integer nonlinear} optimization. 
\begin{mdframed}[backgroundcolor=gray!10, roundcorner=10pt, 
  innerleftmargin=7pt, innerrightmargin=7pt, 
  innertopmargin=10pt, innerbottommargin=10pt]
\small
\textbf{Unconstrained Reformulation}
\begin{equation}\label{eq:miblp}
\begin{aligned}
\min_{\bar{\theta}_{k:T-1}} \quad 
& \sum_{t=k}^{T-1} \sum_{\tau=k}^t \Big( \prod_{s=k}^{t} \bar\theta_{s|k}\Big) g_{t,\tau}
    + \lambda \sum_{t=k}^{T-1}(1 - \bar{\theta}_{t|k})
\end{aligned}
\end{equation}
\end{mdframed}
Although we have reduced the dimension of the optimization problem to $(T-k-1)$, 
independent of the state dimension $n$, the formulation in \eqref{eq:miblp} 
remains a MINP, which is still challenging to solve in practice. 
To overcome this difficulty, we further reformulate the problem as a MILP 
without introducing any suboptimality. We begin by defining unified binary monomials 
that capture the switching effects of the scheduling decisions:

{\small
\begin{align}
\mu_{t,\tau} &:= \prod_{s=\tau}^{t} \bar\theta_{s|k}, 
\quad \mbox{\small $\tau=k,\ldots,t;\; t=k,\ldots,T\!-\!1$}.
 \label{eq:def-mu}
\end{align}
}To enforce the binary product definition in~\eqref{eq:def-mu}, 
we impose the following linear inequalities for all {\small$s \in \{\tau,\ldots,t\}$}:

{\small\begin{equation}
\begin{aligned}
 \sum_{s=\tau}^{t} \bar\theta_{s|k} - (t-\tau)~ \le~  \mu_{t,\tau} ~&\le~ \bar\theta_{s|k}, 
\end{aligned}
\end{equation}
}These constraints guarantee that $\mu_{t,\tau}=1$ if and only if 
$\bar\theta_{s|k}=1$ for all $s \in [\tau,t]$; otherwise $\mu_{t,\tau}=0$. 
This construction is a special case of the McCormick relaxation~\cite{mccormick1976computability}, 
which reduces to an \emph{exact} formulation when applied to binary variables. 
Consequently, the nonlinear product in~\eqref{eq:def-mu} can be replaced with its linear counterpart, 
leading to the following MILP reformulation of the objective:
\begin{mdframed}[backgroundcolor=gray!10, roundcorner=10pt,
  innerleftmargin=2pt, innerrightmargin=2pt,
  innertopmargin=2pt, innerbottommargin=2pt]
\small
\textbf{~~MILP Reformulation}
{\small
\begin{align}
\begin{split} \label{eq:milp}
\min_{\bar{\theta}_{k:T-1},\,\{\mu_{t,\tau}\}} \quad & 
\sum_{t=k}^{T-1} \!\sum_{\tau=k}^{t} \mu_{t,\tau}\, g_{t,\tau}
\;+\; \lambda \sum_{t=k}^{T-1} \big(1-\bar\theta_{t|k}\big) \\
\text{s.t.}\quad
& \bar\theta_{t|k} \in \{0,1\}, \quad 
  \mu_{t,\tau} \in \{0,1\}, 
\\
& \mu_{t,\tau} \le \bar\theta_{s|k}, \quad \forall\, s \in \{\tau,\ldots,t\},
 \\
& \mu_{t,\tau} \ge 
    \sum_{s=\tau}^{t} \bar\theta_{s|k} 
      - \big(t-\tau\big), 
\end{split}
\end{align}}
\end{mdframed}
The entire control-communication joint optimization process is presented in Algorithm~\ref{alg:mpc-trigger}. 
\begin{algorithm}
\caption{\small\textsc{Optimal Event-Triggered LQG}}
\label{alg:mpc-trigger}
\begin{algorithmic}[1]
  \Require System matrices {\small$A,B$}; cost weights {\small$Q,Q_T,R$}; 
           process-noise covariance {\small$\Sigma^w$}; communication penalty {\small$\lambda$}; 
           prediction horizon {\small$T$}
  \Ensure Optimal transmission sequence {\small$\Theta^* = \{\theta_0^*, \dots, \theta_{T-1}^*\}$
          and control inputs {\small$\{u_0^*, \dots, u_{T-1}^*\}$}}

  \State Compute control gains {\small$L_k$} and weight matrices {\small$\Gamma_k$} 
         by solving the Riccati recursion for {\small$k=0,\dots,T-1$}
  \State Precompute kernels {\small$G_{t,\tau}$} and scalars 
         {\small$g_{t,\tau} = \mathrm{tr}(\Gamma_t G_{t,\tau})$}
         for all {\small$0 \le \tau \le t \le T-1$}
  \State Initialize: {\small$\Sigma_0^s = (x_0 - \mathbb{E}[x_0])(x_0 - \mathbb{E}[x_0])^\T$}

  \For{{\small$k = 0$} \textbf{to} {\small$T-1$}}
      \If{{\small$e_k^{s\T}\Gamma_k e_k^s \ge \lambda$}} 
          \Comment{{\footnotesize send is optimal (Thm.~\ref{thm:one-step-certificates})}}
          \State {\small$\bar\theta_{k|k} \gets 0$}
      \ElsIf{{\small $e_k^{s\T} W_k e_k^s\! \le\! \lambda$}} 
          \Comment{{\footnotesize skip is optimal (Thm.~\ref{thm:one-step-certificates})}}
          \State {\small$\bar\theta_{k|k} \gets 1$}
      \Else
          \State Solve MILP~\eqref{eq:milp} to obtain 
                 {\small$\{\bar\theta_{t|k}\}_{t=k}^{T-1}$}
      \EndIf

      \If{$\bar\theta_{k|k} = 0$}
          \State Transmit {\small$x_k$}; reset estimation error: {\small$e_k \gets 0$}
          \State Update: {\small$\mathbb{E}[x_k \mid \mathcal{Z}_k] \gets x_k$}
      \Else
          \State Predict: {\small$\mathbb{E}[x_k \mid \mathcal{Z}_k] \gets 
          A\,\mathbb{E}[x_{k-1} \mid \mathcal{Z}_{k-1}] + B\,u_{k-1}$}
      \EndIf

      \State Apply control: {\small$ u_k \gets -L_k\,\mathbb{E}[x_k \mid \mathcal{Z}_k]$}
      \State Update scheduler error covariance: 
             {\small$\Sigma_k^s \gets (x_k - \mathbb{E}[x_k \mid \mathcal{Z}_k])
             (x_k - \mathbb{E}[x_k \mid \mathcal{Z}_k])^\T$}
  \EndFor
\end{algorithmic}
\end{algorithm}

\begin{figure}[t]
  \centering
\includegraphics[width=0.7\linewidth,height=0.5\linewidth, trim={130 280 120 280}, clip]{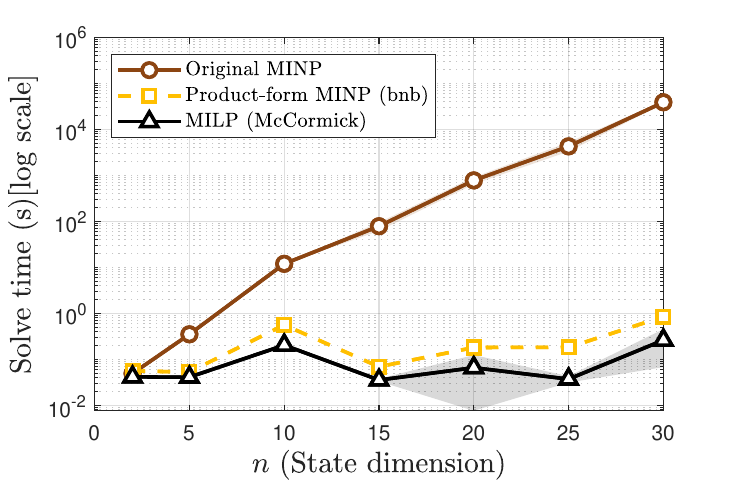}
  \caption{\footnotesize Average optimization time (seconds) vs.\ problem size $n$ on a log~scale for a horizon $T=10$; shaded regions indicate ± one standard deviation across trials (some of them are not visible due to the log scale and small values).}
  \label{fig:time-vs-n}
  \vspace{-4mm}
\end{figure}

While the MILP formulation~\eqref{eq:milp} introduces auxiliary binaries $\mu_{t,\tau}$ (for $k\le \tau\le t\le T-1$) and additional constraints from the (exact) McCormick linearization, this modest size increase is outweighed by a key benefit: products of binaries are replaced by an equivalent linear representation. Consequently, the unconstrained MINP in~\eqref{eq:miblp} becomes a standard MILP with a linear objective and linear constraints. This conversion enables the use of modern MILP solvers (e.g., \textsc{Gurobi}), which are far more efficient than general MINP solvers.
Fig.~\ref{fig:time-vs-n} compares the average runtimes of the three formulations: 
MINP (Matrix Recursion Form) and MILP were solved with \textsc{Gurobi}, 
while the MINP (Unconstrained Reformulation) was solved using YALMIP’s 
nonlinear branch-and-bound solver (\texttt{bnb}). The Matrix Recursion MINP 
scales exponentially, and the Unconstrained Reformulation suffers from 
solver inefficiency\footnote{\texttt{bnb} provides no guarantee of global optimality; due to the problem’s nonconvexity and multiple local minima.} , whereas the MILP achieves nearly constant solution 
times. Across $n=2$-$30$, the reformulation yields speedups of up to $10^5$, verified over multiple randomized trials. All models were implemented in YALMIP~(20230622) on MATLAB; MINP and MILP instances were solved using Gurobi~12.0.1 with default tolerances and no warm-start information, while the product-form MINP used YALMIP’s \texttt{bnb}. Experiments were run on a 3.60GHz Intel Core i7-12700K CPU (12 cores, 32GB RAM).

\section{Efficiency and Optimality of {\small MPC} }  \label{sec:MPC_certificates}

In this section, we present two main technical results. First, we show that the computational complexity of the {\small MPC}  approach can be further reduced by deriving a threshold-based policy to determine whether the optimization at timestep 
$k$ needs to be solved in the first place.
Second, we show that our proposed method outperforms any deterministic policy, including widely used periodic policies. 

\begin{theorem}[One--Step Optimality Certificates]
\label{thm:one-step-certificates}
Consider \eqref{eq:discrete-system}, the error dynamics \eqref{eq:error-dynamics},
and the cost \eqref{eq:cost-function2} with {\small$\Gamma_k:=L_k^\T S_k L_k\succeq0$}.
For {\small$k\in\{0,\ldots,T-1\}$} define the tail Gramian

{\small
\[
W_k:=\sum_{j=0}^{T-1-k}(A^j)^\T\Gamma_{k+j}A^j.
\]}
Let {\small$\mathrm{Ben}_k(e_k^s)$} be the expected cost reduction obtained by sending at time $k$
(i.e., skip minus send). Then

{\small
\begin{equation}
\label{eq:sandwich}
e_k^{s\T}\Gamma_k e_k^s-\lambda
\;\le\;
\mathrm{Ben}_k(e_k^s)
\;\le\;
e_k^{s\T}W_k e_k^s-\lambda .
\end{equation}}
Consequently,
if {\small$e_k^{s\T}\Gamma_k e_k^s\ge\lambda$} then send is optimal, and
if {\small$e_k^{s\T}W_k e_k^s\le\lambda$} then skip is optimal.
Both bounds are tight.
\end{theorem}
\begin{proof}
At time $k$, we define the cost-to-go for each available action, along with their optimal values, as follows.
{\small
\begin{align*}
J_k^{\text{skip}}(\Theta)
&:= \mathbb{E}\!\left[\sum_{t=k}^{T-1}\!\bigl(\|e_t\|_{\Gamma_t}^2+\lambda \theta_t\bigr)\,\middle|\,\theta_k=0,\ e_k^s\right]\\
J_k^{\text{send}}(\Theta)
&:= \mathbb{E}\!\left[\sum_{t=k}^{T-1}\!\bigl(\|e_t\|_{\Gamma_t}^2+\lambda \theta_t\bigr)\,\middle|\,\theta_k=1,\ e_k^s\right]
\end{align*}
\begin{align*}
J_k^{\text{skip},\star}&:=\min_{\Theta}J_k^{\text{skip}}(\Theta)\\
J_k^{\text{send},\star}&:=\min_{\Theta}J_k^{\text{send}}(\Theta)
\end{align*}
}We define the benefit of transmitting as the reduction in cost relative to skipping, i.e.,
{\small
\[
\mathrm{Ben}_k(e_k^s):=J_k^{\text{skip},\star}-J_k^{\text{send},\star}.
\]
}Bounding by evaluating at the opposite policies gives
{\small\begin{equation}
\label{eq:Ben-brackets}
J_k^{\text{skip},\star}-J_k^{\text{send}}(\Theta^{\text{skip}})
\;\le\;
\mathrm{Ben}_k(e_k^s)
\;\le\;
J_k^{\text{skip}}(\Theta^{\text{send}})-J_k^{\text{send},\star},
\end{equation}
}where {\small$\Theta^{\text{skip}}\in\arg\min_\Theta J_k^{\text{skip}}(\Theta)$} and
{\small$\Theta^{\text{send}}\in\arg\min_\Theta J_k^{\text{send}}(\Theta)$}.

At decision time $k$, not transmitting leads to a state-error cost $\|e_k^s\|_{\Gamma_k}^2$, whereas transmitting incurs the fixed communication penalty $\lambda$. Consequently, the instantaneous expected gain from choosing to transmit is
{\small\[
\,e_k^{s\top}\Gamma_k e_k^s-\lambda.
\]
}We can derive a compact expression for the error
{\small\[
e_{k+j}=\mu_{k+j,k+1}\,A^{j}e_k^s+\eta_{k+j} \qquad j\ge 1,
\]
}where 
{\small$\mu_{k+j,k+1}:=\prod_{s=k+1}^{k+j}(1-\theta_s)=1$} indicates no success on $(k,k+j]$, 
and $\eta_{k+j}$ collects process-noise terms (independent of $e_k^s$ and zero-mean):
\[
\eta_{k+j} := \sum_{r=0}^{j-1}\mu_{k+j,k+1+r}\,A^{\,j-1-r} w_{k+r}.
\]
The \emph{difference} of the costs depends only on the deterministic $e_k^s$ term and the indicators $\mu$. Evaluating the right-hand side of~\eqref{eq:Ben-brackets} at $\Theta^{\text{send}}$ yields
{\small\begin{align*}
    J_k^{\text{skip}}(\Theta^{\text{send}})-J_k^{\text{send},\star}
= &\; \,e_k^{s\top}\Gamma_k e_k^s-\lambda\\
  &\; + \sum_{j=1}^{T-1-k}\mu_{k+j,k+1}^{\text{send}}\,
     \|e_k^s\|_{A^{j\!\top}\Gamma_{k+j}A^j}^2 .
\end{align*}
}The worst (largest) tail arises if no success occurs after $k$, i.e.,
$\mu_{k+j,k+1}^{\text{send}}=1$ for all $j$, giving
{\small\[
\mathrm{Ben}_k(e_k^s)\;\le\;
\,e_k^{s\top}\,
\underbracket{\Bigl(\Gamma_k+\sum_{j=1}^{T-1-k}\!A^{j\!\top}\Gamma_{k+j}A^j\Bigr)}_{W_k}
\,e_k^s-\lambda.
\]
}Similarly, evaluating the left-hand side of~\eqref{eq:Ben-brackets} at $\Theta^{\text{skip}}$ gives
{\small\begin{align*}
    J_k^{\text{skip},\star}-J_k^{\text{send}}(\Theta^{\text{skip}})
= &\; \,e_k^{s\top}\Gamma_k e_k^s-\lambda\\
  &\; + \sum_{j=1}^{T-1-k}\mu_{k+j,k+1}^{\text{skip}}\,
     \|e_k^s\|_{A^{j\!\top}\Gamma_{k+j}A^j}^2 .
\end{align*}
}The tightest lower bound occurs if an immediate success follows $k$, i.e.,
{\small$\mu_{k+j,k+1}^{\text{skip}}=0$} for all $j$, which yields
{\small\[
\mathrm{Ben}_k(e_k^s)\ge \,e_k^{s\top}\Gamma_k e_k^s-\lambda.
\]
}Combining the two bounds establishes the theorem.
\end{proof}

\begin{theorem}[{\small MPC}  dominates open-loop deterministic schedules]
\label{thm:mpc-dominates-ol}
Consider \eqref{eq:discrete-system}–\eqref{eq:error-dynamics} with stage cost
$\ell(e_k,\theta_k)=e_k^\top\Gamma_k e_k+\lambda\theta_k$.
Let $\Theta^{\mathrm{det}}$ denote any fixed deterministic schedule.
Let $\Theta^{\mathrm{MPC}}$ be the receding-horizon policy that, at each time $k$,
solves its MILP subproblem to optimality using the current covariance and applies
{\small$\theta^*_{k\mid k}$}. Then, for any {\small$\Theta^{\mathrm{det}}$},

{\small\[
  J(\Theta^{\mathrm{MPC}})\ \le\ J(\Theta^{\mathrm{det}}).
\]}
\end{theorem}
\begin{proof}
Let $\mathcal{I}_k$ be the data available to the scheduler at time $k$. For each $k$, the receding-horizon action $\theta^{*}_{k\mid k}$ is defined as the optimizer of the MILP subproblem at time $k$ conditioned on $\mathcal{I}_k$. Likewise, $\theta^{*}_{k\mid t}$ denotes the action for time $k$ planned at time $t \le k$ (under the subproblem solved at time $t$). We prove the result in two steps.

Step 1 (MPC $\le$ open-loop plan).
We will show

{\small
\begin{equation}\label{eq:MPC-vs-OL}
\underbrace{\mathbb{E}\!\left[\sum_{k=0}^{T-1}\ell\!\big(e_k,\theta^{*}_{k\mid k}\big)\right]}_{\text{Expected performance under MPC}}
\;\le\;
\underbrace{\mathbb{E}\!\left[\sum_{k=0}^{T-1}\ell\!\big(e_k,\theta^{*}_{k\mid 0}\big)\right]}_{\stackrel{\text{expected performance of  open-loop plan }}{\tiny \text{computed at time $t=0$}}}.
\end{equation}
}The proof is by backward induction, using conditional optimality at each time and the tower property of conditional expectation.
\emph{Base case ($k=T-1$).}
By optimality of the receding-horizon decision at time $T-1$,
{\small
\[
\ell\!\big(e_{T-1},\theta^{*}_{T-1\mid T-1}\big)
\;\le\;
\ell\!\big(e_{T-1},\theta^{*}_{T-1\mid T-2}\big)
\quad\text{for all $e^s_{T-1}$\quad a.s.}
\]
}The equality holds only if the decision computed at time $T-2$ (i.e., $\theta^{*}_{T-1\mid T-2}$) is still optimal at time $T-1$.

Taking conditional expectations given $\mathcal{I}_{T-2}$ and using monotonicity of conditional expectation,

{\small
\[
\mathbb{E}\!\left[\ell\!\big(e_{T-1},\theta^{*}_{T-1\mid T-1}\big)\,\middle|\,\mathcal{I}_{T-2}\right]
\;\le\;
\mathbb{E}\!\left[\ell\!\big(e_{T-1},\theta^{*}_{T-1\mid T-2}\big)\,\middle|\,\mathcal{I}_{T-2}\right],
\]
}Applying the tower property yields:
{\small
\[
\mathbb{E}\!\left[\ell\!\big(e_{T-1},\theta^{*}_{T-1\mid T-1}\big)\right]
\;\le\;
\mathbb{E}\!\left[\ell\!\big(e_{T-1},\theta^{*}_{T-1\mid T-2}\big)\right].
\]
}\emph{Inductive step (from $j+1$ down to $j$).}
Fix $j \in \{0, \ldots, T-2\}$ and assume as induction hypothesis that the inequality holds from $j+1$ onward.

{\small
\[
\mathbb{E}\!\left[\sum_{k=j+1}^{T-1}\ell\!\big(e_k,\theta^{*}_{k\mid k}\big)\right]
\;\le\;
\mathbb{E}\!\left[\sum_{k=j+1}^{T-1}\ell\!\big(e_k,\theta^{*}_{k\mid j}\big)\right].
\]
}By construction of the receding-horizon step, the pair {\small$\Big(\theta^{*}_{j\mid j},\,\{\theta^{*}_{k\mid j}\}_{k=j+1}^{T-1}\Big)$} is (by definition) optimal for the one-step decision at $j$ when the future plan $\{\theta^{*}_{k\mid j}\}_{k>j}$ is taken as fixed. Hence, almost surely,

{\small
\begin{equation*}\label{eq:stepj}
\begin{aligned}
&\ell\!\big(e_j,\theta^{*}_{j\mid j}\big) 
   + \mathbb{E}\!\left[\sum_{k=j+1}^{T-1}\ell(e_k,\theta^{*}_{k\mid j}) \,\middle|\,\mathcal{I}_j\right] \\
&\qquad\le\;
\ell\!\big(e_j,\theta^{*}_{j\mid j-1}\big) 
   + \mathbb{E}\!\left[\sum_{k=j+1}^{T-1}\ell(e_k,\theta^{*}_{k\mid j-1}) \,\middle|\,\mathcal{I}_j\right].
\end{aligned}
\end{equation*}
}Taking unconditional expectations and applying the tower property to both conditional terms gives:

{\small
\begin{equation*}\label{eq:stepj}
\begin{aligned}
&\mathbb{E}\!\left[\ell\!\big(e_j,\theta^{*}_{j\mid j}\big)\right]
 + \mathbb{E}\!\left[\sum_{k=j+1}^{T-1}\ell(e_k,\theta^{*}_{k\mid j})\right] \\
&\qquad\le\;
\mathbb{E}\!\left[\ell\!\big(e_j,\theta^{*}_{j\mid j-1}\big)\right]
 + \mathbb{E}\!\left[\sum_{k=j+1}^{T-1}\ell(e_k,\theta^{*}_{k\mid j-1})\right].
\end{aligned}
\end{equation*}
}Combining this with the induction hypothesis yields:

{\small
\[
\mathbb{E}\!\left[\sum_{k=j}^{T-1}\ell\!\big(e_k,\theta^{*}_{k\mid k}\big)\right]
\;\le\;
\mathbb{E}\!\left[\sum_{k=j}^{T-1}\ell\!\big(e_k,\theta^{*}_{k\mid j-1}\big)\right].
\]
}Plugging $j =  1$ and then adding 
$\mathbb{E}\!\left[\ell\!\big(e_0,\theta^{*}_{0\mid 0}\big)\right]$ to both sides
yields \eqref{eq:MPC-vs-OL}.

Step 2 (open-loop plan $\preceq$ any deterministic schedule).
By definition, the open-loop sequence $\{\theta^{*}_{k\mid 0}\}_{k=0}^{T-1}$ minimizes {\small $\mathbb{E}\!\left[\sum_{k=0}^{T-1}\ell\!\big(e_k,\theta_k\big)\middle| \mathcal{I}_0\right]$} over all admissible deterministic schedules fixed at time $0$. Any deterministic schedule $\Theta^{\mathrm{det}}=\{\theta^{\mathrm{det}}_{k\mid 0}\}_{k=0}^{T-1}$ is feasible for this problem; therefore,

{\small
\[
\mathbb{E}\!\left[\sum_{k=0}^{T-1}\ell\!\big(e_k,\theta^{*}_{k\mid 0}\big)\right]
\;\le\;
\mathbb{E}\!\left[\sum_{k=0}^{T-1}\ell\!\big(e_k,\theta^{\mathrm{det}}_{k\mid 0}\big)\right].
\]
}Together, these two steps establish the theorem.

\end{proof}
Step~1 of the proof shows that, in expectation, the receding-horizon {\small MPC} strategy 
is at least as good as the offline open-loop plan 
(the full-horizon MILP solved once at time~$0$). 
In fact, the open-loop plan can produce the same performance as the MPC \textit{only if} the open-loop plan is optimal at every sub-optimization problem (i.e., the optimization from stage $k$ onward).
If there exists a time $k$ such that the solution to the MPC problem is unique and $\theta^*_{k|k} \ne \theta^*_{k|k-1}$, the MPC outperforms the openloop policy. 
Step~2 shows that, in expectation, the offline open-loop plan 
is at least as good as any deterministic fixed schedule.

\section{Case Study: Double Integrator } \label{sec:case-study}
We evaluate the proposed MILP-based event-triggered scheduling on a double integrator, modeling a robot moving along a line with position–velocity dynamics:

{\small
\[
x_{k+1} =
\begin{bmatrix}
1 & T_s \\[2pt]
0 & 1
\end{bmatrix} x_k +
\begin{bmatrix}
\frac{T_s^2}{2} \\[2pt]
T_s
\end{bmatrix} u_k + w_k,
\]
}where {\small\(x_k = [p_k \; v_k]^\T\)} is the state (position and velocity), {\small\(u_k\)} is the control input (acceleration), and {\small\(w_k\)} is zero-mean noise with covariance \(\Sigma^w\). The sampling period is set to {\small\(T_s = 0.1\,\text{s}\)}. Simulation parameters are:

{\small
\[
T = 25,  \quad \Sigma^w = 0.5 I, \quad \lambda = 100.
\]}
\begin{enumerate}[leftmargin=*]
    \item \textbf{Open-Loop Prediction:} No transmissions;
          the controller relies solely on open-loop predictions. 
    \item \textbf{Continuous Communication:} State transmitted at every step, i.e., $\theta_k=1$ for all $k$ (ideal baseline). 
    \item \textbf{Offline Schedule:} Fixed transmission schedule computed once 
          from the full-horizon MILP. 
    \item \textbf{{\small MPC} Scheduler:} Receding-horizon MILP solved at each step using 
          the current error covariance. 
\end{enumerate}

This comparison illustrates the trade-off between communication and control performance across baseline, idealized, and optimized strategies.

\begin{figure}
    \centering
    \begin{subfigure}[b]{0.48\columnwidth}
        \centering
        \includegraphics[width=\linewidth,height=0.6\linewidth, trim={100 240 130 280}, clip]{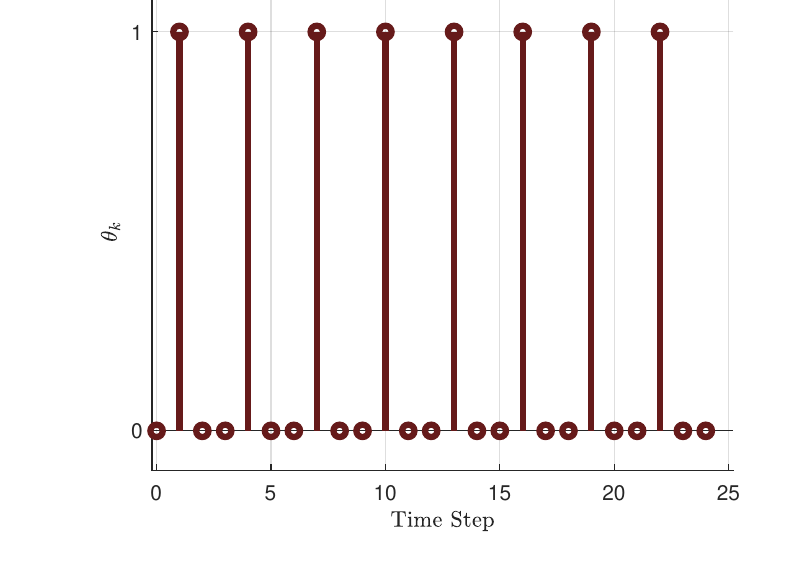}
        \caption{Offline transmission schedule.}
        \label{fig:schedule-offline}
    \end{subfigure}
    \hfill
    \begin{subfigure}[b]{0.48\columnwidth}
        \centering
        \includegraphics[width=\linewidth,height=0.6\linewidth, trim={100 240 130 280}, clip]{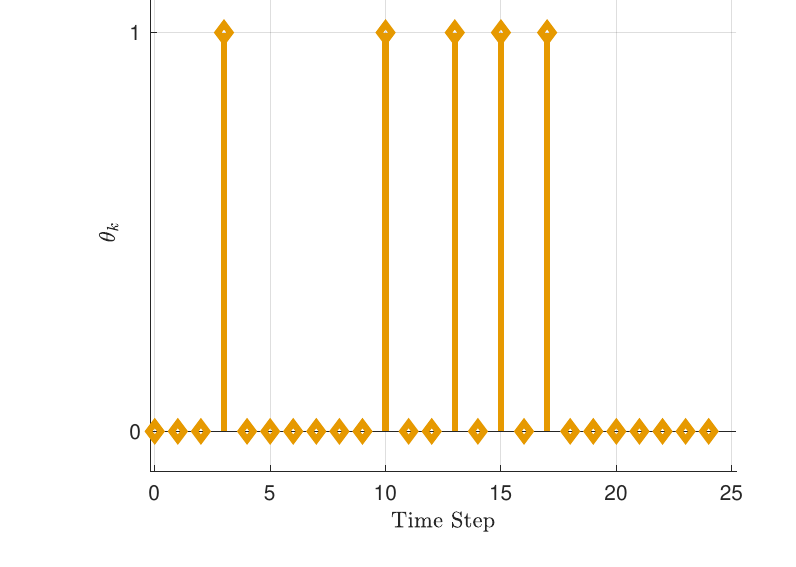}
        \caption{MPC transmission schedule.}
        \label{fig:schedule-mpc}
    \end{subfigure}

    \caption{\footnotesize Transmission schedules under offline vs.\ MPC strategies for a representative disturbance realization.}
    \label{fig:schedule-comparison}
\end{figure}

Figure~\ref{fig:schedule-comparison} compares the transmission schedules of the offline and {\small MPC}  strategies for a representative disturbance realization. The offline method follows a fixed  pattern, transmitting at regular intervals regardless of system behavior or estimation error. In contrast, the {\small MPC}  strategy adapts to real-time uncertainty, triggering transmissions only when necessary. This adaptivity enables MPC to substantially reduce communication without sacrificing performance.

\begin{figure}
    \centering
    \begin{subfigure}[b]{0.48\columnwidth}
        \centering
        \includegraphics[width=\linewidth, trim={135 280 130 280}, clip]{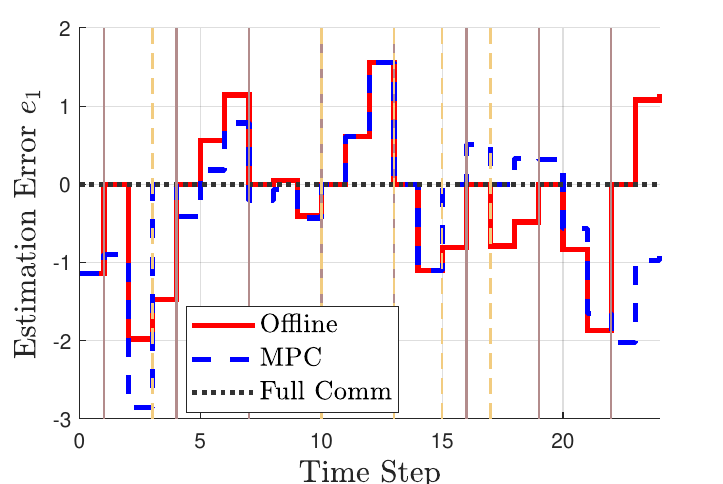}
        \caption{\footnotesize $x(1)$ estimation error.}
        \label{fig:esterror-offline}
    \end{subfigure}
    \hfill
    \begin{subfigure}[b]{0.48\columnwidth}
        \centering
        \includegraphics[width=\linewidth, trim={135 280 130 280}, clip]{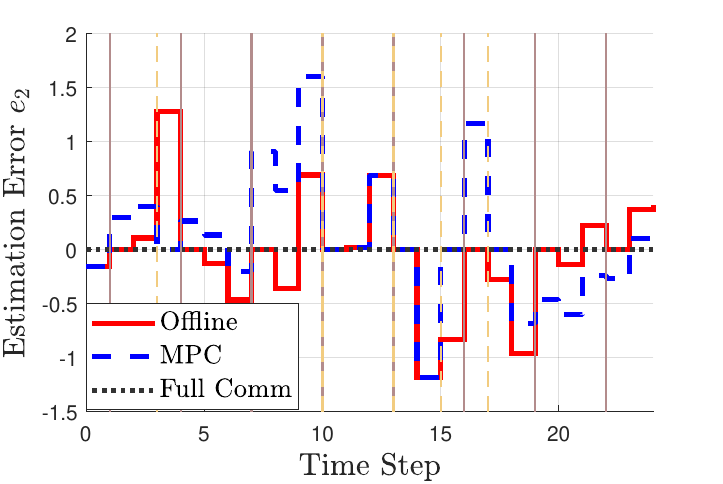}
        \caption{\footnotesize $x(2)$ estimation error.}
        \label{fig:esterror-mpc}
    \end{subfigure}
    \caption{\footnotesize 
    Estimation error trajectories under offline vs.\ MPC scheduling. The lighter shade vertical lines represent the scheduling time which are the same as the ones in Fig.~\ref{fig:schedule-comparison}.}
    \label{fig:esterror-comparison} \vspace{-12 pt}
\end{figure}

\begin{figure}
    \centering
    \begin{subfigure}[b]{0.48\columnwidth}
        \centering
        \includegraphics[width=\linewidth, trim={150 270 150 270}, clip]{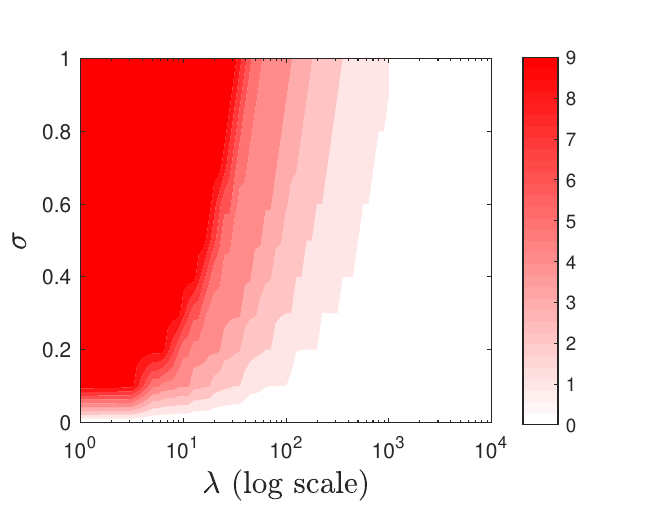}
        \caption{\footnotesize Offline optimization.}
        \label{fig:numcom-offline}
    \end{subfigure}
    \hfill
    \begin{subfigure}[b]{0.48\columnwidth}
        \centering
        \includegraphics[width=\linewidth,trim={150 270 150 270}, clip]{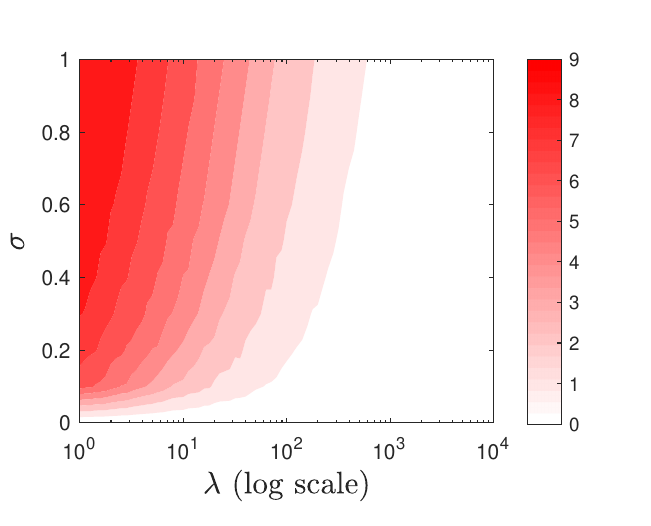}
        \caption{\footnotesize MPC scheduling.}
        \label{fig:numcom-mpc}
    \end{subfigure}
    \caption{\footnotesize Average communication count under offline vs.\ MPC scheduling across different values of $\lambda$ and noise covariance.}
    \label{fig:comm-count-comparison}\vspace{-8 pt}
\end{figure}

Figure~\ref{fig:esterror-comparison} shows the estimation error trajectories under the three scheduling strategies. Both offline and {\small MPC} reset the error to zero upon transmission. The offline policy follows a fixed schedule, making it insensitive to disturbance realizations, whereas the {\small MPC}  policy adapts to real-time uncertainty and transmits only when needed. As a result, {\small MPC}  achieves error bounds comparable to the offline schedule in both $x(1)$ and $x(2)$ while using fewer transmissions. By contrast, full communication removes estimation error but requires transmission at every step.
The efficiency of {\small MPC} in balancing communication and control/estimation is illustrated in Figure~\ref{fig:comm-count-comparison}, which reports contours of the \emph{average} number of transmissions as a function of the communication penalty {\small$\lambda$} and the noise covariance scale {\small$\sigma$} (with {\small$\Sigma^w=\sigma I$}). Both approaches depend on $\sigma$ because the offline schedule is designed using {\small$\Sigma^w$}; however, the offline policy is open-loop—once {\small$(\lambda,\sigma)$} are fixed it yields a static transmission pattern that does not react to the particular noise \emph{realization} along a trajectory. In contrast, {\small MPC}  re-optimizes online using the realized estimation error/innovation and therefore adapts its transmission frequency to both cost and uncertainty; this adaptive effect is most pronounced for low to moderate values of {\small$\lambda$}, leading to more efficient use of communication.

\begin{figure}[t]
    \centering
    \begin{subfigure}[b]{0.48\columnwidth}
        \centering
        \includegraphics[width=\linewidth, trim={150 270 150 270}, clip]{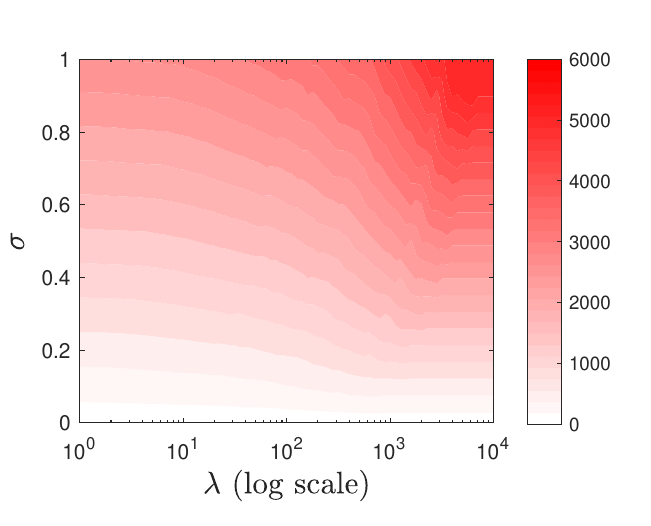}
        \caption{\footnotesize MPC scheduling.}
        \label{fig:avgJ-mpc}
    \end{subfigure}
    \hfill
    \begin{subfigure}[b]{0.48\columnwidth}
        \centering
        \includegraphics[width=\linewidth, trim={150 270 150 270}, clip]{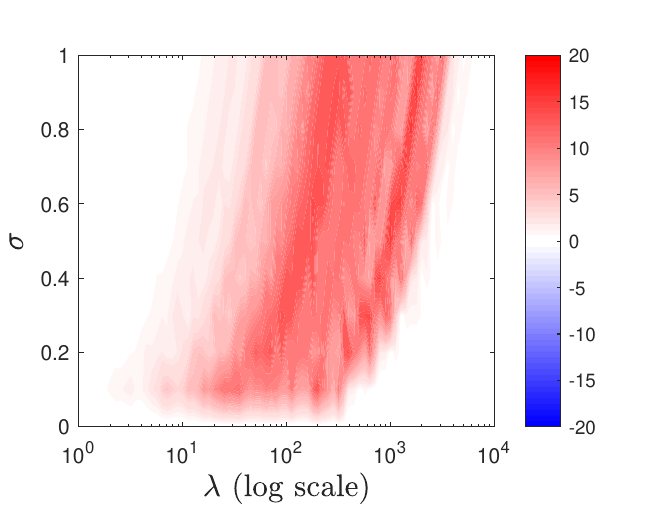}
        \caption{\footnotesize Improvement over Offline.}
        \label{fig:avgJ-offline}
    \end{subfigure}
    \caption{\footnotesize Average total cost under MPC and percentage improvement over offline scheduling across different values of $\lambda$ and noise covariance $\sigma$}.
    \label{fig:avg-cost-comparison}
\end{figure}

A similar pattern appears in the cost landscape of Figure~\ref{fig:avg-cost-comparison}. The {\small MPC}  strategy consistently achieves lower total cost than the offline baseline, with the advantage most pronounced when communication is expensive and transmissions are limited. These results underscore the value of incorporating real-time information into both control and scheduling decisions.

To assess robustness, we ran Monte Carlo simulations over {\small1000} disturbance realizations. Table~\ref{tab:cost-comparison-detailed} summarizes the results. 
{\small MPC}  achieves the lowest average cost of {\small5694.34} with only on average of {\small4.46} transmissions per run, about half that of offline {\small(5893.16, 8.14)}. 
Full communication eliminates error but at the highest cost {\small(6956.29, 25.0)}, while prediction-only minimizes transmissions {\small(1.0)} but incurs the largest cost {\small(17860.73)}.
{\small
\begin{table}[t]
\centering
\caption{\small Average cost and communication count under different strategies.}
\label{tab:cost-comparison-detailed}
\begin{tabular}{|l|c|c|}
\hline
\textbf{Strategy} & \textbf{Cost} & \textbf{Comm. Avg.} \\
\hline
\multicolumn{3}{|c|}{\textbf{Single Case Study}} \\
\hline
Offline Schedule           & 3420.23    & 8.00 \\
{\small MPC}  Scheduler               & 3046.57    & 5.00 \\
Continuous Communication& 5107.83    & 25.00 \\
Open-Loop Prediction        & 5988.35    & 0.00 \\
\hline
\multicolumn{3}{|c|}{\textbf{Monte Carlo Average (1000 Seeds)}} \\
\hline
Offline Schedule           & 5893.16    & 8.14 \\
{\small MPC}  Scheduler               & 5694.34    & 4.46 \\
Continuous Communication& 6956.29    & 25.00 \\
Open-Loop Prediction        & 17860.73   & 0.00 \\
\hline
\end{tabular}
\end{table}}

\begin{figure}
    \centering 
    \includegraphics[width=0.6\columnwidth, trim={140 250 150 265}, clip]{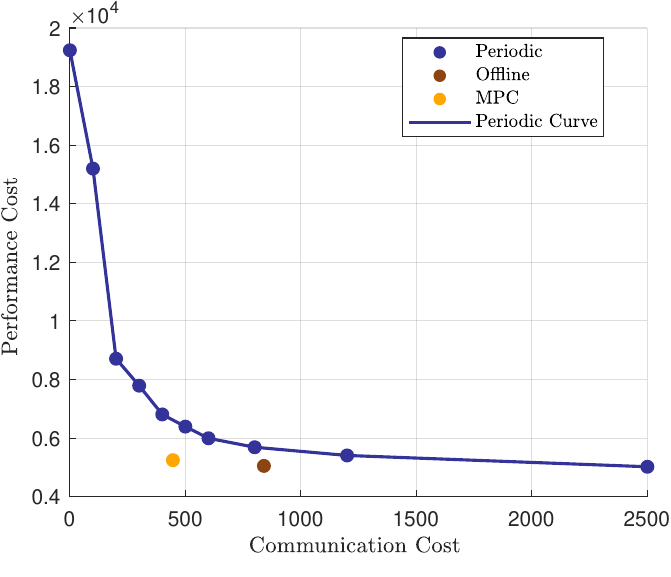}
   \caption{\footnotesize Comparison of control strategies with the baseline periodic strategy. The \textit{Performance Cost} is {\footnotesize $ \mathbb{E}\!\left[
\sum_{k=0}^{T-1} 
(\|x_k\|_Q^2 + \|u_k\|_R^2)
+ \|x_T\|_{Q_T}^2
\right]
$} and the \textit{Communication Cost} is {\footnotesize $\mathbb{E}[\sum_{k=0}^{T-1}\lambda\theta_k]$}.}
    \label{fig:periodically}
\end{figure}

Figure~\ref{fig:periodically} compares control strategies in terms of the trade-off between communication and performance cost. The periodic baseline (dark blue points and fitted curve) shows that higher communication frequency improves performance. In contrast, both offline optimization (brown) and {\small MPC}  scheduling (yellow) achieve lower costs than the periodic baseline at comparable or reduced communication levels. This demonstrates the advantage of optimized and adaptive scheduling over fixed periodic triggering in reducing communication while maintaining performance. 

Overall, the {\small MPC} approach provides a scalable, efficient solution for resource-aware estimation and control, well suited to bandwidth- or energy-constrained systems.

\section{Conclusion} \label{sec:conclusions}
We introduced a linear reformulation for event-triggered scheduling in stochastic networked control systems, converting a challenging nonlinear optimization problem into a tractable MILP. This reformulation uncovered structural insights into how estimation error, communication decisions, and control performance interact, and led to the development of intuitive transmission certificates. Our numerical results show that the method dramatically reduces computational effort while achieving a better trade-off between control cost and communication usage compared to standard approaches. These results point to the potential of this framework as a scalable foundation for more complex applications, such as multi-agent coordination and large-scale networked systems.

\bibliographystyle{ieeetr}
\bibliography{main}

@article{aggarwal2025interq,
  title={InterQ: A {DQN} Framework for Optimal Intermittent Control},
  author={Aggarwal, Shubham and Maity, Dipankar and Ba{\c{s}}ar, Tamer},
  journal={IEEE Control Systems Letters},
  year={2025},
  publisher={IEEE}
}

@article{maity2021optimal,
  title={Optimal controller synthesis and dynamic quantizer switching for {Linear-Quadratic-Gaussian systems}},
  author={Maity, Dipankar and Tsiotras, Panagiotis},
  journal={IEEE Transactions on Automatic Control},
  volume={67},
  number={1},
  pages={382-389},
  year={2021},
  publisher={IEEE}
}

@inproceedings{molin2009lqg,
  title={On {LQG} joint optimal scheduling and control under communication constraints},
  author={Molin, Adam and Hirche, Sandra},
  booktitle={Proceedings of the 48h IEEE Conference on Decision and Control (CDC) held jointly with 2009 28th Chinese Control Conference},
  pages={5832-5838},
  year={2009},
  organization={IEEE}
}

@article{mccormick1976computability,
  title={Computability of global solutions to factorable nonconvex programs: Part {I}---Convex underestimating problems},
  author={McCormick, Garth P},
  journal={Mathematical Programming},
  volume={10},
  number={1},
  pages={147--175},
  year={1976},
  publisher={Springer}
}

@article{molin2012optimality,
  title={On the optimality of certainty equivalence for event-triggered control systems},
  author={Molin, Adam and Hirche, Sandra},
  journal={IEEE Transactions on Automatic Control},
  volume={58},
  number={2},
  pages={470--474},
  year={2012},
  publisher={IEEE}
}

@article{soleymani2021value,
  title={Value of information in feedback control: Quantification},
  author={Soleymani, Touraj and Baras, John S and Hirche, Sandra},
  journal={IEEE Transactions on Automatic Control},
  volume={67},
  number={7},
  pages={3730--3737},
  year={2021},
  publisher={IEEE}
}

@article{soleymani2022value,
  title={Value of information in feedback control: Global optimality},
  author={Soleymani, Touraj and Baras, John S and Hirche, Sandra and Johansson, Karl H},
  journal={IEEE Transactions on Automatic Control},
  volume={68},
  number={6},
  pages={3641--3647},
  year={2022},
  publisher={IEEE}
}

@phdthesis{thelander2020lqg,
  title={On LQG-Optimal Event-Based Sampling},
  author={Thelander Andr{\'e}n, Marcus},
  year={2020},
  school={Lund University}
}

@article{heemels2012introduction,
  title={An introduction to event-triggered and self-triggered control},
  author={Heemels, WPMH and Johansson, KH and Tabuada, P},
  journaltitle={IEEE Control Systems Magazine},
  volume={32},
  number={5},
  pages={50--65},
  year={2012}
}

@article{kostina2017rate,
  title   = {Rate--cost tradeoffs in control},
  author  = {Kostina, Victoria and Hassibi, Babak},
  journal = {IEEE Transactions on Automatic Control},
  volume  = {63},
  number  = {11},
  pages   = {3390--3404},
  year    = {2018}
}

@article{imer2010optimal,
  title        = {Optimal estimation with limited measurements},
  journaltitle = {International Journal of Systems, Control and Communications},
  volume       = {2},
  number       = {1-3},
  pages        = {5--29},
  year         = {2010},
  author       = {Imer, O. C. and Ba{\c{s}}ar, T.}
}

@article{rabi2012adaptive,
  title        = {Adaptive sampling for linear state estimation},
  journal = {SIAM Journal on Control and Optimization},
  volume       = {50},
  number       = {2},
  pages        = {672--702},
  year         = {2012},
  author       = {Rabi, M. and Moustakides, G. V. and Baras, J. S.}
}

@article{lipsa2011remote,
  title        = {Remote state estimation with communication costs for first-order {LTI} systems},
  journal = {IEEE Transactions on Automatic Control},
  volume       = {56},
  number       = {9},
  pages        = {2013--2025},
  year         = {2011},
  author       = {Lipsa, G. M. and Martins, N. C.}
}

@article{maity2020minimal,
  title={Minimal feedback optimal control of linear-quadratic-{Gaussian} systems: No communication is also a communication},
  author={Maity, Dipankar and Baras, John S},
  journal={IFAC-PapersOnLine},
  volume={53},
  number={2},
  pages={2201--2207},
  year={2020},
  publisher={Elsevier}
}

@article{mamduhi2025network,
  title={Network-Aware Optimal Sampling for Stochastic Control Systems over Dynamic Networks},
  author={Mamduhi, Mohammad H and Maity, Dipankar},
  journal={IEEE Control Systems Letters},
  year={2025},
  publisher={IEEE},
  volume ={9},
  pages={1808--1813}
}

@article{maity2019optimal,
  title={Optimal event-triggered control of nondeterministic linear systems},
  author={Maity, Dipankar and Baras, John S},
  journal={IEEE Transactions on Automatic Control},
  volume={65},
  number={2},
  pages={604--619},
  year={2019},
  publisher={IEEE}
}

@incollection{aastrom1997fundamental,
  title={Fundamental limitations of control system performance},
  author={{\AA}str{\"o}m, Karl Johan},
  booktitle={Communications, Computation, Control, and Signal Processing: a tribute to Thomas Kailath},
  pages={355--363},
  year={1997},
  publisher={Springer}
}

@inproceedings{maity2015eventa,
  title={Event based control of stochastic linear systems},
  author={Maity, Dipankar and Baras, John S},
  booktitle={2015 International Conference on Event-based Control, Communication, and Signal Processing (EBCCSP)},
  pages={1--8},
  year={2015},
  organization={IEEE}
}

@article{afshari2024communication,
  title={Communication-and Control-Aware Optimal Quantizer Selection for Multi-Agent Control},
  author={Afshari, Mohammad and Maity, Dipankar and Tsiotras, Panagiotis},
  journal={IEEE Control Systems Letters},
  volume={8},
  pages={2385--2390},
  year={2024},
  publisher={IEEE}
}

@article{maity2023regret,
  title={Regret-Optimal Cross-Layer Co-Design in Networked Control Systems—Part {II: Gauss-Markov} Case},
  author={Maity, Dipankar and Mamduhi, Mohammad H and Lygeros, John and Johansson, Karl H},
  journal={IEEE Communications Letters},
  volume={27},
  number={11},
  pages={2879--2883},
  year={2023},
  publisher={IEEE}
}

@article{mamduhi2023regret,
  title={Regret-optimal cross-layer co-design in networked control systems—Part {I}: General case},
  author={Mamduhi, Mohammad H and Maity, Dipankar and Johansson, Karl H and Lygeros, John},
  journal={IEEE Communications Letters},
  volume={27},
  number={11},
  pages={2874--2878},
  year={2023},
  publisher={IEEE}
}

@article{maity2018optimal,
  title={Optimal {LQG} control under delay-dependent costly information},
  author={Maity, Dipankar and Mamduhi, Mohammad H and Hirche, Sandra and Johansson, Karl Henrik and Baras, John S},
  journal={IEEE control systems letters},
  volume={3},
  number={1},
  pages={102--107},
  year={2018},
  publisher={IEEE}
}

@article{maity2021optimalb,
  title={Optimal {LQG} control of networked systems under traffic-correlated delay and dropout},
  author={Maity, Dipankar and Mamduhi, Mohammad H and Hirche, Sandra and Johansson, Karl H},
  journal={IEEE Control Systems Letters},
  volume={6},
  pages={1280--1285},
  year={2021},
  publisher={IEEE}
}

@inproceedings{hashemi2016energy,
  title={Energy hub management by using decentralized robust model predictive control},
  author={Hashemi, Zahra and Ramezani, Amin and Moghaddam, Mohsen Parsa},
  booktitle={2016 4th international conference on control, instrumentation, and automation (ICCIA)},
  pages={105--110},
  year={2016},
  organization={IEEE}
}

\clearpage 
\newpage

\end{document}